\documentclass[9pt,twocolumn,twoside]{osajnl}
\pdfoutput=1
\usepackage{graphicx}
\graphicspath{{pictures/}}
\DeclareGraphicsExtensions{.pdf,.png,.jpg,.eps}

\journal{pr} 

\setboolean{shortarticle}{true}

\title{Short spin waves detection by means of the magneto-optical intensity effect}

\author[1,2,*]{Olga V. Borovkova}
\author[2]{Savelii V. Lutsenko}
\author[1,2]{Andrey N. Kalish}

\affil[1]{Russian Quantum Center, Novaya str. 100, Skolkovo, Moscow Region, Russia, 143025}
\affil[2]{Faculty of Physics, M.V. Lomonosov Moscow State University, Leninskie Gory, Moscow, Russia, 119991}

\affil[*]{Corresponding author: o.borovkova@rqc.ru}

\begin{abstract}
It is proposed a novel method of the spectrally selective detection of the short spin waves (or magnons) by means of the transverse magneto-optical (MO) intensity effect in transmission in the magnetoplasmonic nanostructure. The method is based on the analysis of the MO effect spectrum versus the modulation of the sample magnetization (i.e. spin wave) and related spatial symmetry breaking in the magnetic layer. The spatial symmetry breaking leads to the appearance of MO effect modulation at the normal incidence of light in the spectral range of the optical states (SPP and waveguide modes) and the breaking of the antisymmetry of the effect with respect to the sign of the incidence angle of light. Besides it is revealed that the magnitude of the MO effect varies with respect to the phase shift between the spin wave and the plasmonic grating as a harmonic function with a period equal to the magnon wavelength. All these facts allow to detect the spin waves of the certain wavelength propagating in the nanostructure by measuring of the MO effect in the nanostructure.
\end{abstract}

\setboolean{displaycopyright}{true}

\begin{document}

\maketitle

Recently, an interest in spin waves, or magnons, raised both from the side of fundamental science and from the point of view of practical applications due to a variety of the possibilities that are offered by magnonics for the data storage and processing systems \cite{Barman:2021, Satoh:2012, Kalashnikova:2007, Nikitov:2015, Kimel:2005, Kampen:2002, Kirilyuk:2010}. Magnonic devices would have much lower power consumption and operate at frequencies up to dozens of THz. But nowadays the novel materials and magnetic nanostructures, where the excitation, control and detection of the spin waves can be realized, are required. In recent years the optical techniques for magnonics have been actively developing. The optomagnonics provides several approaches for spin waves generation \cite{Kimel:2005, Kampen:2002, Kirilyuk:2010, Hansteen:2005, Hansteen:2006}. However, the up-to-date methods of magnons detection \cite{Kalashnikova:2007, Kirilyuk:2010, Kimel:2005, Atoneche:2010, Savochkin:2017, Yoshimin:2017, Chernov:2018},  have a number of limitations. One of the main problems is that the existing methods detect spin waves in a wide frequency range. To solve this problem, in this Letter it is proposed to employ the magnetoplasmonic nanostructures, that make it possible to detect spin waves in a narrow frequency range by means of the magneto-optical intensity effect in transmission enhanced nearby the optical resonances related to the excitation of the surface plasmon polaritons and waveguide modes.
\newline
The magnetic nanostructures are well-known for their ability to enhance the magneto-optical effects or to amplify and enrich their properties. \cite{Wurtz:2008, Belotelov:2011, Baryshev:2013, Armelles:2013, Borovkova:2019, Kreilkamp:2013, Chekhov:2014}. For instance, recently it was shown that the spatial symmetry breaking in the magnetoplasmonic nanostructures can lead to the novel MO effect, the transverse intensity magneto-optical effect in transmission \cite{Borovkova:2020}. One of the features of the reported phenomenon was the fact that MO effect is nonzero at the normal incidence of light. In \cite{Borovkova:2020} the spatial symmetry breaking was created due to the asymmetry of the unit cell geometry in the plasmonic grating. The excitation of the spin wave, or the waves of magnetization in a continuous magnetic medium, also acts as a spatial symmetry breaking in the nanostructure. However, the MO intensity effect \textit{does} turn to zero under certain conditions despite the spatial symmetry breaking made by the spin wave. Here we address the case when the period of the plasmonic grating is equal to the integer number of the magnetization oscillations period (or periods of the spin wave), the spatial symmetry becomes broken inside one grating cell. In this Letter we analyze how the magnetization modulation influences the MO intensity effect in transmission nearby the spectral range of the SPP and waveguide modes. Based on this it is proposed the novel approach how to detect the short spin waves of the narrow range of the wavelength by means of the measurements of the MO effect in the nanostructure.

\begin{figure}[h]
\centering
\includegraphics[width=0.9\linewidth]{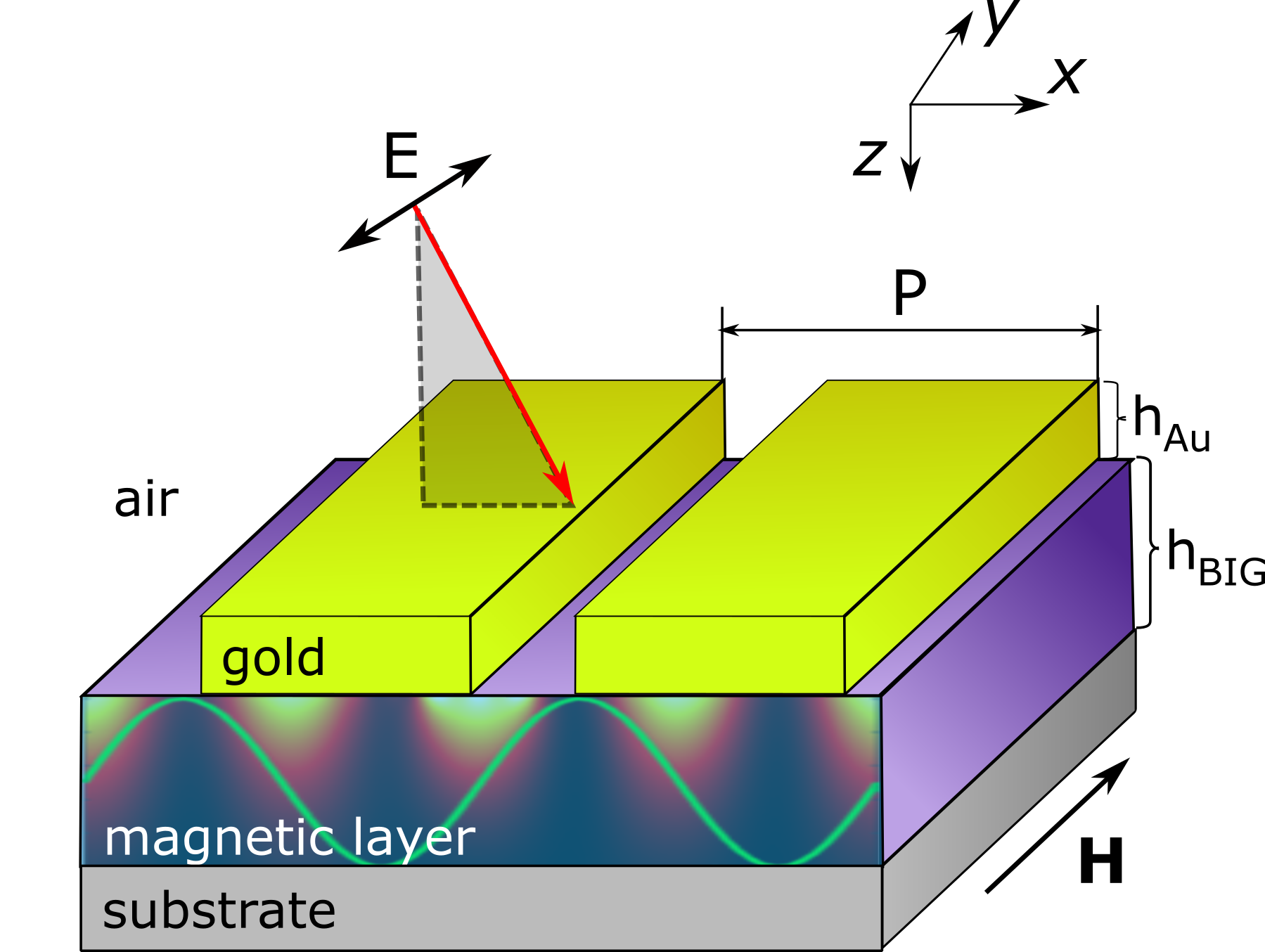}
\caption{The scheme of the addressed magnetoplasmonic nanostructure. A modulation of the magnetization inside the magnetic layer is shown by the green line. The input plane $p$-polarized light is shown schematically.}
\label{fig:Scheme}
\end{figure}

We address the magnetoplasmonic nanostructures composed of a thin layer (or a film) of a magnetic material deposited on the surface of a nonmagnetic dielectric substrate of gadolinium gallium garnet (GGG). The top surface of the magnetic layer is covered by a one-dimensional plasmonic grating made of metal. A ferrimagnetic dielectric of bismuth-substituted iron garnet is chosen as a magnetic layer due to its transparency in the visible spectral range and the significant MO response \cite{Chernov:2018, Geller:1963, Borovkova2018, Levy:19}. We assume that the constant external magnetic field $\bf{H}$ is applied along the $y$-axis and the magnetization saturation of the ferrimagnetic film is achieved. The magnons propagate in the plane of the ferrimagnetic film perpendicular to the plasmonic grating axis. The direction of a modulation coincides with the direction of modulation of a one-dimensional plasmonic grating (see Fig. \ref{fig:Scheme}). We describe the magnetization modulation in the ferrimagnetic layer as a modulation of the local gyration of the material
\begin{equation}
g(x) = g_0 sin(\frac{2\pi}{\lambda_{SW}}x+\phi),
\label{eq:gyration}
\end{equation}
where $g_0$ is the amplitude of the magnetization modulation, it corresponds to the saturation magnetization of the ferrimagnetic layer, $\lambda_{SW}$ is the wavelength of the spin wave, and $\phi$ is a phase shift between the spin wave and the plasmonic grating. The magnetization distribution in the sample given by Eq. (\ref{eq:gyration}) can be created by exciting of spin waves or magnons. If the period of the spin wave corresponds to the period of the plasmonic grating the MO effects observed in such nanostructure have some peculiarities that can be used for the detection of the magnons.

The nanostucture is illuminated by the plane wave of \textbf{p}-polarized light as it is shown in the Fig. \ref{fig:Scheme}. We vary the wavelength and the incidence angle of the input light and analyze the transmission spectrum of the structure at the different directions of the external magnetic field $\bf{H}$.

The transverse magneto-optical intensity effect in transmission \cite{Borovkova:2020} is determined as a relative change of the transmitted light intensity $T(\bf{M})$ when the structure is re-magnetized
\begin{equation}
\delta_T = 2\frac{T(\bf{M}) - T(\bf{-M})}{T(\bf{M}) + T(\bf{-M})}.
\label{eq:TMOKE}
\end{equation}

Here $\bf{M}$ is the magnetization saturation of the magnetic material emerging when the external magnetic field $\bf{H}$ is applied.

If one considers the \textbf{p}-polarized light falling down the magnetic material with spatial modulation of the gyration (\ref{eq:gyration}), it is easy to show that the relative change of the transmitted light \cite{MOBook, Belotelov:2009} is proportional to the gyration $g$ of the magnetic material. For the considered system it means that

\begin{equation}
\delta_T \approx sin(\frac{2\pi}{\lambda_{SW}}x+\phi).
\label{eq:TMOKE_phi}
\end{equation}

In absence of the plasmonic grating for the input plane wave the integral over the \textit{x}-coordinate results in the opposing impact from positive and negative half periods of the sine function and the phase $\phi)$ doesn't play any role. However, in case of the magnetoplasmonic nanostructure like in the Fig. \ref{fig:Scheme} two aspects become crucial.
\begin{enumerate}
    \item  The coincidence of the period of the plasmonic nanostructure and the magnetization modulation. If the periods are different the MO response inside different periods of the plasmonic grating would be different and they can interfere both constructively and destructively, distorting the response from the entire structure as a whole. So, we limit our consideration by $\lambda_{SW}=P$, the wavelength of the spin wave is equal to the period of the plasmonic nanostructure.

    \item The phase shift $\phi$ determines the spatial displacement between the magnetization modulation and the plasmonic grating, and, ultimately, the (see Eq. (\ref{eq:TMOKE_phi})) the MO effect. So, twice per period, for $\phi=0$ and $\phi=\pi$ the magneto-optical effect in transmission turns to be 0, although it is nonzero for all other cases.
    
\end{enumerate}

Thus, the magnetoplasmonic nanostructure with spatial symmetry breaking can have zero MO effect under the certain conditions. In details the dependence on the phase shift would be discussed further.

To clear up the impact of the spin waves on the MO effects in the magnetoplasmonic nanostructure we analyze the wavelength and angular-resolved transmission and $\delta_T$ spectra of the addressed nanostructure with the spin waves of different periods and compare the results with the case of the uniform magnetization of the ferrimagnetic film. There were considered two set of parameters of the plasmonic grating. The first one was chosen to provide the excitation of the surface plasmon polariton (SPP) modes that occurs in the spectral range of 650-900nm. And the second configuration supports the excitation of the waveguide modes in the magnetic layer at wavelength of 550-625nm. 

For the excitation of the SPP modes the period of the plasmonic grating was chosen to be 630nm and the width of the air gap is 85nm, so $86.5\%$ of the surface is covered by gold. The thickness of the metal is 80nm, the thickness of the ferrimagnetic film is 100nm. Note that such thin ferrimagnetic layer prevent the excitation of the waveguide modes and for them a bit different configuration was chosen.

\begin{figure}[h]
\centering
\includegraphics[width=0.9\linewidth]{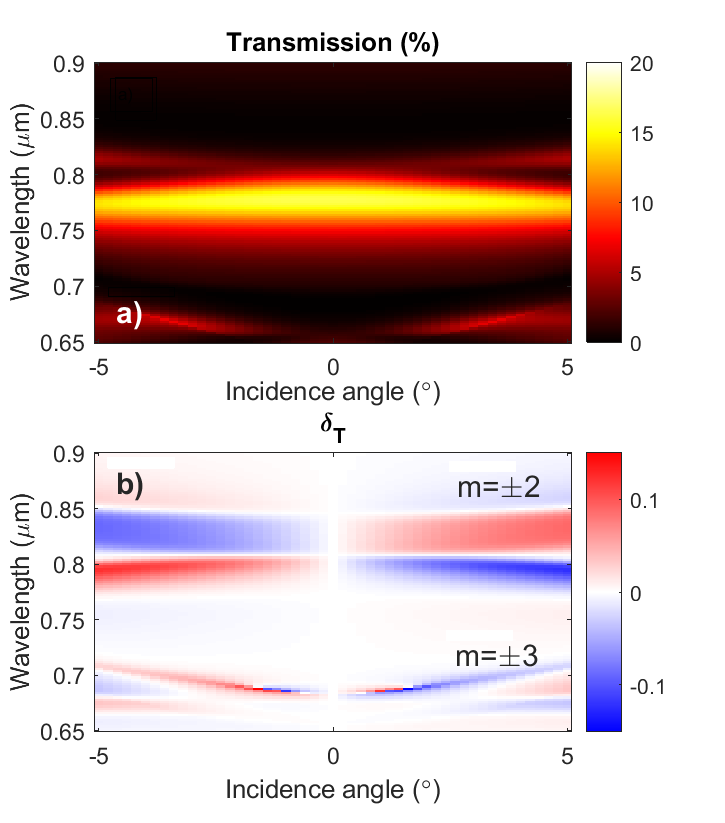}
\caption{Angular and wavelength-resolved transmission (a) and $\delta_T$ (b) spectra of the addressed nanostructure with the uniform spatial distribution of the magnetization in the ferrimagnetic layer. The orders of the SPP modes are denoted by $m$.}
\label{fig:TandDeltaUniform}
\end{figure}

Fig. \ref{fig:TandDeltaUniform} shows the transmission and $\delta_T$ spectra of the nanostructure with uniform magnetization of the ferrimagnetic layer. The typical resonant peculiarities related to the excitation of the surface plasmon polariton (SPP) mode with the orders $m=\pm2$ (upper branch) and $m=\pm3$ (lower branch) are denoted in the Fig. \ref{fig:TandDeltaUniform}. 
The transmission spectrum is symmetric, and the spectrum of the magneto-optical effect is antisymmetric due to the change of input light sign and the corresponding change of the spatial symmetry conditions. Besides that, due to the symmetry reasons the $\delta_T$ spectrum is equal to zero at the normal incidence of light.  

\begin{figure}[h!]
\center{\includegraphics[width=0.9\linewidth]{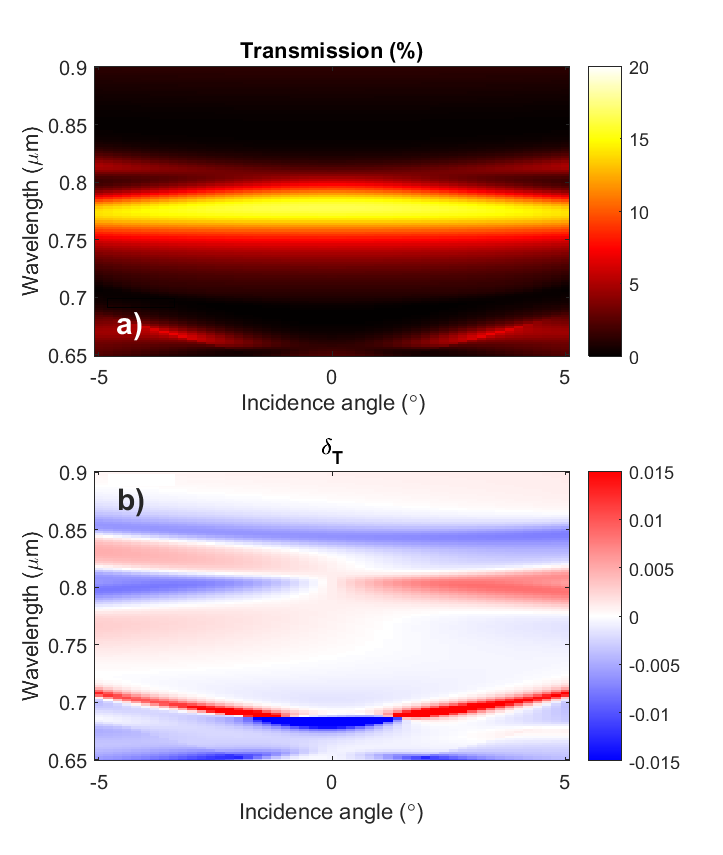}}
\caption{Angular and wavelength-resolved transmission (a) and $\delta_T$ (b) spectra of the addressed nanostructure with the spatial modulation of the magnetization in the ferrimagnetic layer. The magnetization modulation period is equal to the period of the plasmonic grating. The phase shift between the plasmonic grating and the magnetization modulation is 0.}
\label{fig:TandDeltak1}
\end{figure}

When the magnetization of the ferrimagnetic film is modulated due to the excitation of the spin waves, it causes the change in both transmission and $\delta_T$ spectra. In the Fig. \ref{fig:TandDeltak1} the corresponding optical and magneto-optical spectra are given for the cases when the magnetization modulation period is equal to the period of the plasmonic grating. The phase shift was chosen to provide the maximum value of the $\delta_T$.

One can see that in the spectral range of the SPP modes of the order $m=\pm2$ (in particular, near $\lambda=850nm$) the MO effect has a constant sign and its value is almost the same over the wide range of the incident angles. So, the MO effect is nonzero at the normal incidence of light in the spectral range of the SPP modes of the order $m=\pm2$. The SPP modes of the order $m=\pm3$ (near $\lambda=700nm$) also reveal the nonzero MO effect at the normal incidence of light. Therefore, the spatial symmetry breaking caused by the magnetization modulation makes the changes in the MO effect offering the method to detect the presence of the spin wave. The only difficulty is that the magnitude of the $\delta_T$ is 10 times less than for the case when the magnetization is unmodulated (see Fig. \ref{fig:TandDeltaUniform}b).  

\begin{figure}[h]
\center{\includegraphics[width=0.9\linewidth]{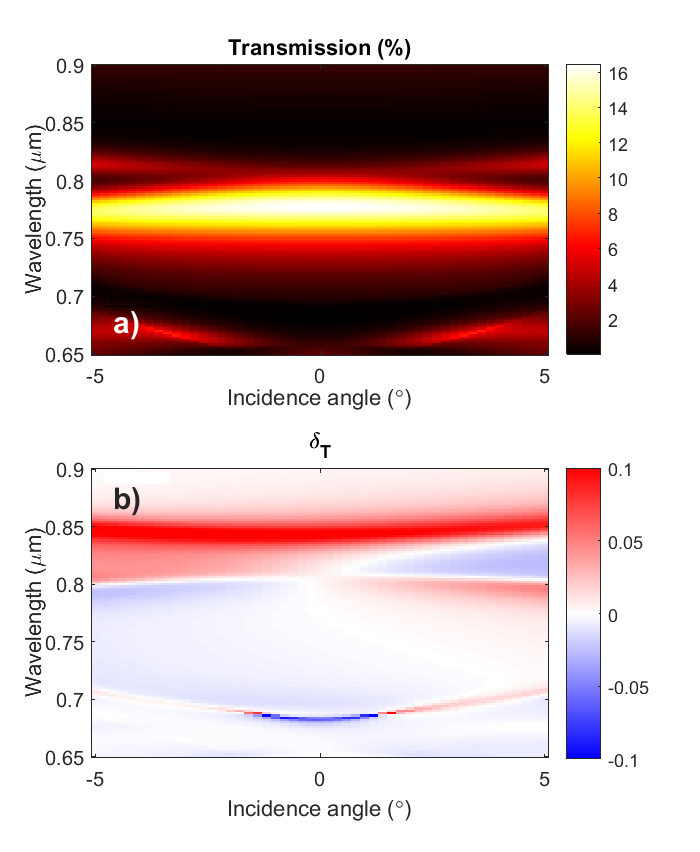}}
\caption{Angular and wavelength-resolved transmission (a) and $\delta_T$ (b) spectra of the addressed nanostructure with the spatial modulation of the magnetization in the ferrimagnetic layer. The magnetization modulation period is equal to the half of the plasmonic grating period. The phase shift between the plasmonic grating and the magnetization modulation is 0.}
\label{fig:TandDeltak2}
\end{figure}

This problem can be overcome when the period of the magnetization modulation is equal to the half of the plasmonic grating period. The reason is that in the considered setting at the interface of the [plasmonic grating]/[ferrimagnetic dielectric] for $\lambda=850nm$ in the air the SPP wave wavelength is about 313nm \cite{Belotelov:2011}. So, when $\lambda_{SW} = P/2$, the SPP mode wavelength $\lambda_{SPP}$ is equal to the $\lambda_{SW}$ and the impact of the spatial symmetry breaking created by the magnetization modulation is stronger. The corresponding transmission and $\delta_T$ spectra are given in the Fig. \ref{fig:TandDeltak2} for the case of zero phase shift. The magnitude of the MO effect is comparable with the case of uniform magnetization distribution.

\begin{figure}[h]
\begin{minipage}[h]{0.5\linewidth}
\center{\includegraphics[width=\linewidth]{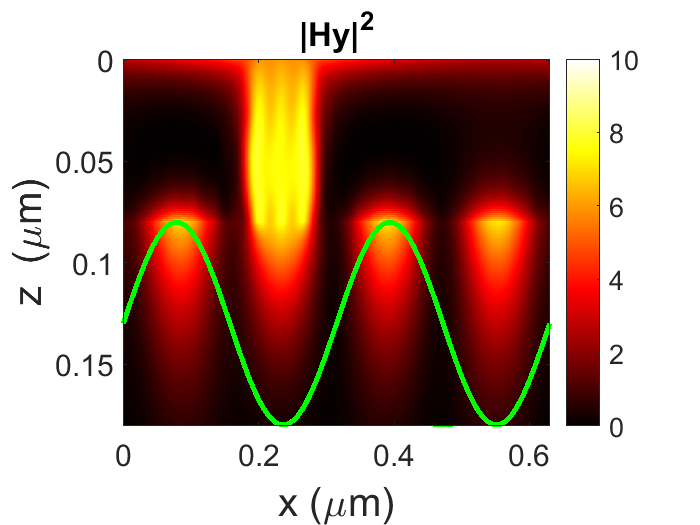}\\ a)}
\end{minipage}
\hfill
\begin{minipage}[h]{0.5\linewidth}
\center{\includegraphics[width=\linewidth]{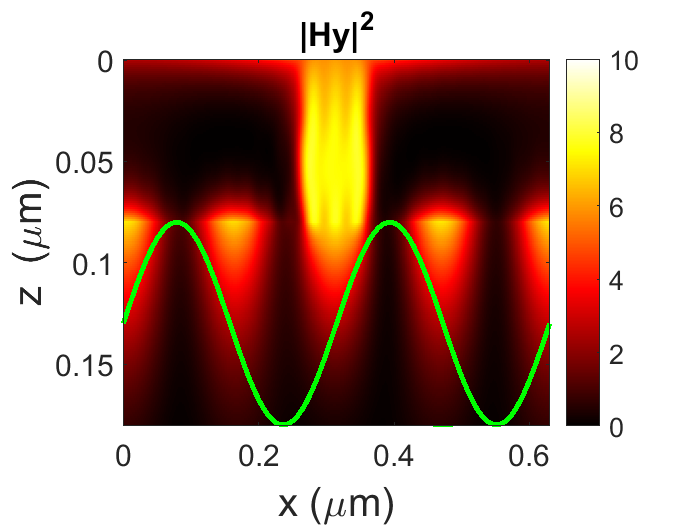}\\ b)}
\end{minipage}
\caption{The spatial distributions of the $|H_y|^2$ and the magnetization modulation (green line) for two different values of phase shift between the plasmonic grating and the magnetization modulation (a) $\phi=1.9rad$ ($\delta_T=0$) and (b) $\phi=2.7rad$ ($\delta_T=0.15$). The magnon wavelength is a half of the plasmonic grating period.}
\label{fig:Fieldk2}
\end{figure}

Now, let's analyze the dependence of the MO effect $\delta_T$ on the phase shift and how it can help to detect the spin waves in the magnetic layer. In the Fig. \ref{fig:Fieldk2} there are given the spatial distributions of the $|H_y|^2$ component of the excited SPP wave and the magnetization modulation with two different phase values. In the Fig. \ref{fig:Fieldk2}a the spatial distribution of the magnetization is non-symmetric inside the plasmonic grating cell, but the impacts of the half periods of the magnetization to the magneto-optical response compensate each other. It makes the whole nanostructure insensitive to the re-magnetization, and although the SPP modes of order $m=\pm2$ are excited in the nanostructure the corresponding resonances do not emerge in the $\delta_T$ spectrum (see Fig. \ref{fig:TandDeltak2}b). On the contrary, the space shift between the the plasmonic grating and the magnetization modulation given in Fig. \ref{fig:Fieldk2}b breaks the spatial symmetry and provides the magneto-optical response for the same spin wave length. These two cases provide different MO effect value. We propose to employ the harmonic dependence of the MO effect on the magnetization modulation for the purposes of spin wave detection. 

\begin{figure}[h]
\begin{minipage}[h]{0.5\linewidth}
\center{\includegraphics[width=\linewidth]{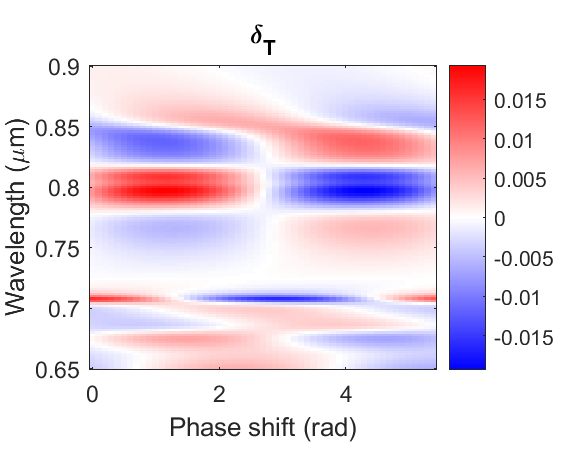}\\ a)}
\end{minipage}
\hfill
\begin{minipage}[h]{0.5\linewidth}
\center{\includegraphics[width=\linewidth]{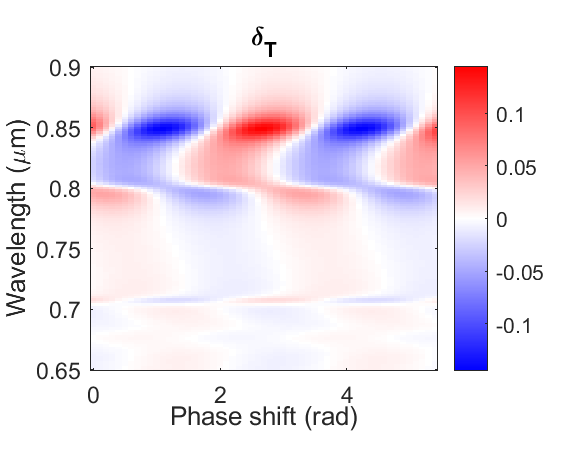}\\ b)}
\end{minipage}
\caption{The magneto-optical effect $\delta_T$ versus the phase shift $\phi$ between the plasmonic grating and the magnetization modulation.}
\label{fig:TMOKE_var_x1}
\end{figure}

Varying the the phase shift between the plasmonic grating and the magnetization modulation we can break and restore the spatial symmetry and modulate the magneto-optical effect. In the Fig. \ref{fig:TMOKE_var_x1} there are given the spectra of the MO effect versus the phase shift for two different wavelength of the spin wave. The dependence of the $\delta_T$ on the phase shift between the plasmonic grating and the magnetization modulation $\phi$ has the period equal to the period of the spin wave wavelength. This fact opens up the novel possibilities to control the magnitude of the effect and, on the other hand, to detect the spin waves in the magnetic layer of the nanostructure. We are not limited by the precise coincidence of the period of the plasmon grating and the wavelength of the spin waves. On the contrary, the higher frequencies of the magnons can be easily detected and distinguished by measuring how the $\delta_T$ spectrum depends on time (i.e. a phase shift).

Besides the SPP modes, the proposed method can be implemented by the excitation of the waveguide modes in the magnetic layer. For this purpose a bit different design of the magnetoplasmonic nanostructure was taken. The period of the plasmonic grating was chosen to be 580nm and the width of the air gap is 75nm. The thickness of the metal is 80nm, the thickness of the ferrimagnetic film was taken greater, 150nm. The thicker layer of the magnetic material is required for the effective excitation of the waveguide modes \cite{Chekhov:2014}. 

\begin{figure}[h]
\begin{minipage}[h]{0.5\linewidth}
\center{\includegraphics[width=\linewidth]{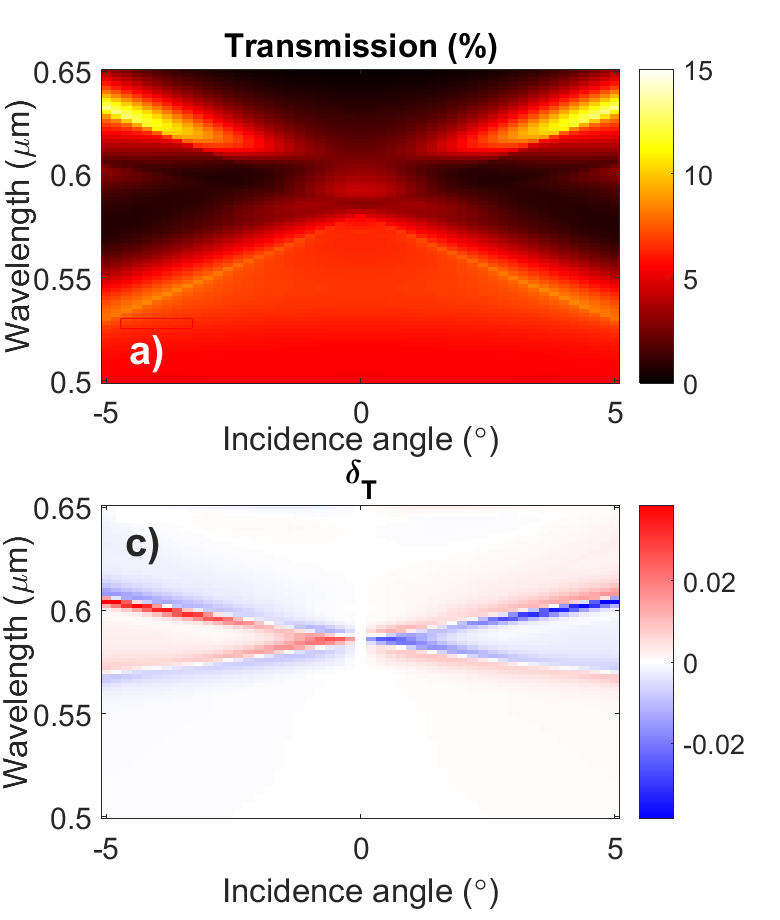}}
\end{minipage}
\hfill
\begin{minipage}[h]{0.5\linewidth}
\center{\includegraphics[width=\linewidth]{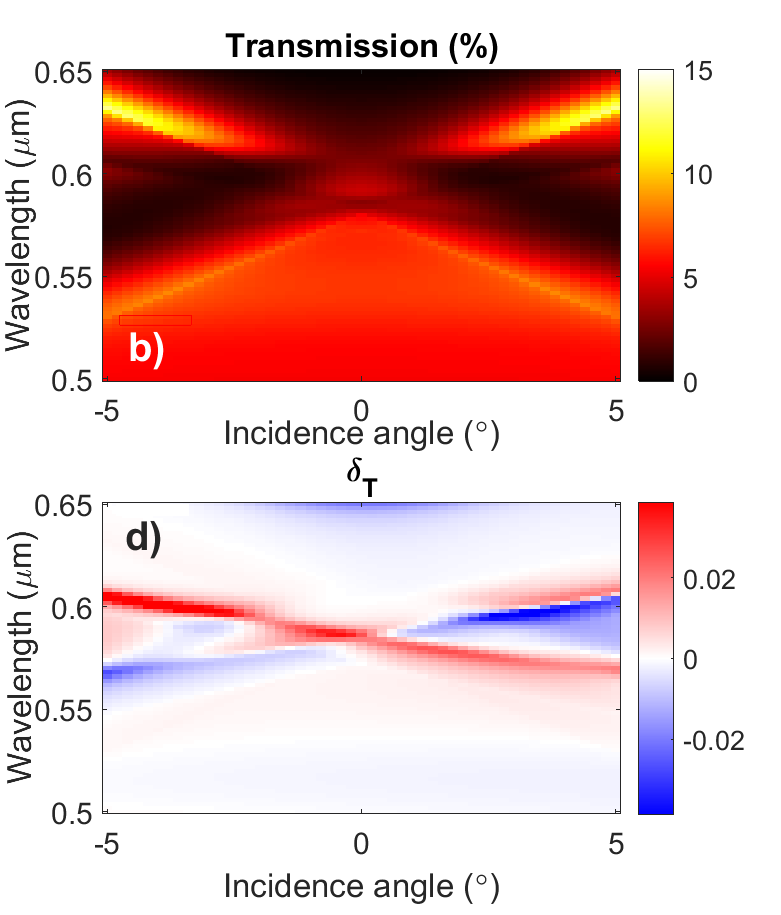}}
\end{minipage}
\caption{Angular and wavelength-resolved transmission (upper plots) and $\delta_T$ (lower plots) spectra of the nanostructure near the waveguide modes excitation area with (right column) and without (left column) the spatial modulation of the magnetization in the ferrimagnetic layer. The phase shift between the plasmonic grating and the magnetization modulation is 0.}
\label{fig:TMOKE_WMs}
\end{figure}

The comparison of the transmission and the magneto-optical effect $\delta_T$ spectra in the nanostructures with the uniform and modulated magnetization is given in the Fig. \ref{fig:TMOKE_WMs}. The right column depicts the results for the magnetization modulation equal to $P/2$, i.e. to the half of the period of the plasmonic grating. 

Similarly to the SPP modes addressed above the difference of the transmission spectra is negligible, but the magneto-optical effect allows one to detect unambiguously the presence of the excited spin waves. The typical nonzero magneto-optical effect accompanied with the similar sign of the $\delta_T$ for opposite incidence angles of the input light is observed.

The spin wave detection in the spectral range of the waveguide modes acts in the same way as the SPP modes do but it is preferrable for the thick magnetic films where the SPP modes are distorted by the multiple waveguide modes. As soon as the same magnetoplasmonic nanostructure sample can support the excitation of both types of modes the frequency range of the spin waves that can be felt by the proposed method is rather wide. Moreover, one sample could contain the set of plasmonic gratings with various parameters, namely, period and the air gap, so scanning the magneto-optical effect from the different gratings one can detect the wide range of the spin waves.
\newline

To sum up, the magnetization modulation created by the spin waves can be detected by means of the magnetoplasmonic nanostructures. It is shown that in presence of the spin waves the transverse MO intensity effect in transmission experiences a harmonic modulation and turns to zero with the periodicity equal to the period of the spin wave. This result can be used for the selective detection of the spin waves presence in the magnetic nanostructure. The proposed approach can be extended also to the near-IR spectral range as well as to the other magneto-optical effects.

\begin{backmatter}
\bmsection{Funding} This study was supported by Russian Science Foundation (project no. 20-72-10159).

\bmsection{Acknowledgments} The authors thank M.A. Kozhaev for fruitful discussions and valuable comments.

\bmsection{Disclosures} The authors declare no conflicts of interest.


\end{backmatter}


\bibliography{main}

\bibliographyfullrefs{main}

\end{document}